\documentclass{iopart}
\usepackage[utf8]{inputenc}
\usepackage{graphicx}
\usepackage{cite}
\usepackage[none]{hyphenat}
\usepackage{alphabeta}
\usepackage{textcomp}
\usepackage{gensymb}

\newcommand{\ddt}[1]{\frac{\mathrm{d}#1}{\mathrm{d}t}}
\newcommand{\Tp}{T_\mathrm{p}}
\newcommand{\degC}{\degree\mathrm{C}}

\begin{document}

\title[ ]{On unipolar and bipolar HiPIMS pulse configurations to enhance energy flux to insulating surfaces}

\author{M. Farahani$^1$, T. Kozák$^1$, A.D. Pajdarová$^1$, T. T\"{o}lg$^1$, J. Čapek$^1$}
\address{$^1$ Department of Physics and NTIS -- European Centre of Excellence, University of West Bohemia, Univerzitní 8, 301 00 Plzeň, Czech Republic}
\ead{kozakt@kfy.zcu.cz }

\begin{abstract}
High-power impulse magnetron sputtering (HiPIMS) delivers a high target power in short pulses, enhancing the ionization and energy of sputtered atoms and providing thus more possibilities to control the film properties. This study explores the effect of various pulse configurations (unipolar HiPIMS, bipolar HiPIMS, chopped unipolar, and chopped bipolar HiPIMS)ease energy flux to an insulated surface (e.g., substrate or growing film). The chopped bipolar HiPIMS configuration, featuring several short positive pulses replacing a single long positive pulse, is introduced, and the total energy fluxes are subsequently measured using a passive thermal probe. Moreover, the effect of the probe's capacitance with respect to the ground is systematically investigated by connecting an external capacitor. Results show that for an insulated surface with low capacitance, bipolar pulse configurations do not significantly increase energy flux to the surface due to its rapid charging by plasma ions. Conversely, high surface capacitance facilitates an increase in energy flux, as a large potential difference between the plasma and the surface remains even for a long positive pulse. For medium surface capacitance (tens of nF), chopping the positive pulse in bipolar HiPIMS effectively increases the energy delivered to the film by discharging the surface in the off-times. The thermal probe measurements also confirm that energy to the film can be increased for unipolar HiPIMS configurations by splitting the negative pulse into several shorter pulses.
\end{abstract}

\noindent{\it Keywords\/}: High power impulse magnetron sputtering, Bipolar HiPIMS, Chopped HiPIMS, Insulating surfaces, Energy flux, Thermal probe

\submitto{\PSST}

\maketitle

High-Power Impulse Magnetron Sputtering (HiPIMS) is an advanced technique used for thin film deposition, where the target material is sputtered within short, high-power pulses. These conditions produce a high fraction of ionized species and broaden their ion energy distribution functions, offering greater flexibility in tailoring the properties of the films. \cite{Lundin2019, Zeman2017, Kumar2020, Gudmundsson2012, Batkova2020}. 

The energy of the ions may be further enhanced as they travel through a region where the plasma potential decreases. This occurs in the plasma sheath near the chamber surfaces (including substrates), where the plasma potential drops from a value corresponding to the conditions in the plasma volume (typically a few volts) to zero for grounded surfaces or to negative values for surfaces at floating potential. The extra energy gained by the ions is then determined by the difference between the plasma potential in the volume and at the surface.

Additional ion acceleration is necessary if the energy delivered to the growing film by the ions is insufficient for the desired film properties. For conductive films and substrates, effective ion acceleration can be achieved through conventional HiPIMS with a continuous DC bias. The negative bias lowers the surface potential of the growing film, enlarging the potential difference responsible for the ion acceleration. However, this approach may result in the unintended incorporation of Ar gas into the growing film \cite{Viloan2021, Villamayor2018}. To solve these issues, synchronized pulsed DC bias is often preferred. By timing the bias to coincide with the metal-ion-rich portion of the HiPIMS pulse, this method minimizes gas entrapment in the material\cite{REZEK2023, Villamayor2018}.

Bipolar HiPIMS has been proposed as an alternative when a DC substrate bias voltage is ineffective, such as for insulating surfaces. This approach introduces a positive voltage following the negative pulse, also establishing the potential difference for the ion acceleration. In this case, the potential difference arises because the plasma potential in the plasma volume rises to a value comparable to the applied positive voltage during the positive pulse \cite{Hippler2020, Law2023}. However, for insulating surfaces, this acceleration quickly diminishes as the surface charges by impacting ions to the floating potential, which is close to the elevated potential of the plasma volume \cite{Batkova2020, Viloan2021}. Therefore, this approach also provides limited control over ion energy. So, a more effective solution is needed to improve the usability of this technique for insulating surfaces. 

Building upon the developments in standard HiPIMS, chopping techniques have been successfully introduced to further enhance the performance of the process \cite{Barker2014, Hnilica2023, Fekete2017}. By dividing each pulse into smaller segments, "chopped HiPIMS'' (or multi-pulse HiPIMS, abbreviated m-HiPIMS) has demonstrated significant improvements in ionization efficiency and deposition rate, which surpass the results of conventional HiPIMS \cite{Barker2014,Antonin2015}. Given these promising outcomes, applying the chopping technique to bipolar HiPIMS offers an exciting opportunity to further optimize the deposition processes. 

In this study, it is proposed to use several short segments of negative and positive pulses instead of using a single long negative and positive pulse to enhance energy flux to insulating surfaces/substrates. The effectiveness of this chopped bipolar HiPIMS technique is evaluated mainly based on the measured energy fluxes to a static thermal probe by comparing six pulse configurations, including standard unipolar and bipolar HiPIMS, as well as several configurations with multiple negative and positive pulses. We show that for certain conditions the chopped bipolar HiPIMS may provide a significant increase in the energy flux to an insulating surface.

\begin{figure}
    \centering
    \includegraphics[width=85mm]{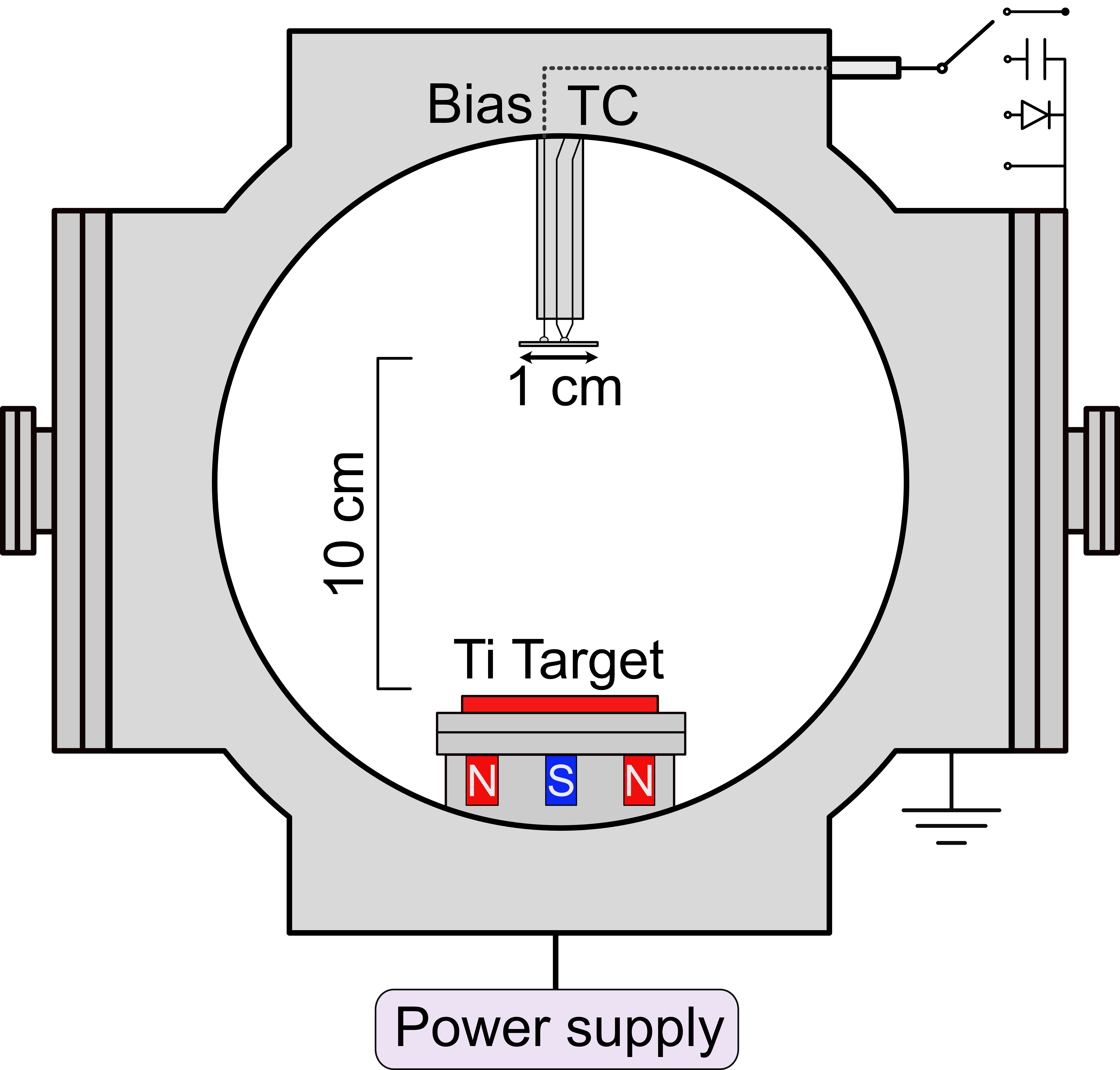}
       \caption{Experimental setup with the passive thermal probe. TC denoted the connection of the integrated thermocouple used to measure the probe's temperature. The various connections of the biasing wire are shown schematically.}
    \label{fig:system}
\end{figure}

\begin{figure*}[t]
    \centering
    \includegraphics[width=1\linewidth]{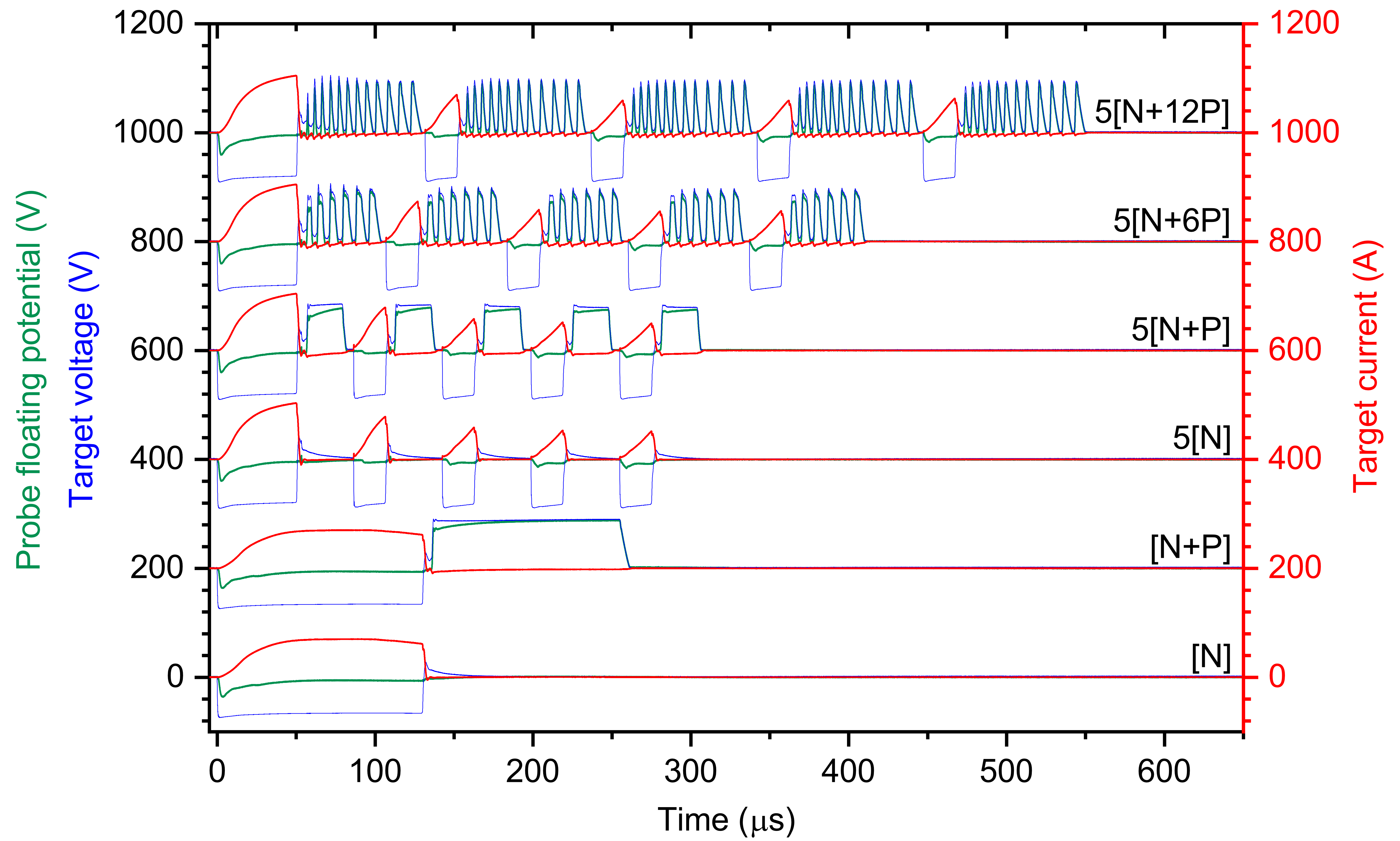}
       \caption{Waveforms of the floating potential of the thermal probe, the target voltage and the target current for the investigated pulse configurations. It is important to note that the target voltage is reduced by a factor of 10 during negative pulses to enable a compact presentation of the waveforms while maintaining good resolution of the voltages during the positive pulses.}
    \label{fig:waveforms}
\end{figure*}

The experimental setup utilized a magnetron (VT100, Gencoa) with an unbalanced (type II) magnetic field configuration and a planar titanium target (100~mm in diameter) housed within a stainless-steel vacuum chamber with a volume of 24.4 liters. A turbomolecular pump, supported by a scroll pump, was used to evacuate the chamber (\fref{fig:system}). Before the experiments, the base pressure inside the chamber was reduced below 4$\times10^{-4}\,\mathrm{Pa}$. The magnetron was driven by a custom-built pulsing unit capable of generating bipolar pulses, controlled by a waveform generator (Rigol DG4102), and powered with two DC power supplies (GS10, ADL for negative pulses and GEN600-1.3, TDK-Lambda for positive pulses). The target current was measured by a current probe (TCP303, Tektronix), and the target and thermal probe (see below) voltages were measured by two voltage probes (TESTEC, 100x attenuation); all connected to a digital oscilloscope (Picoscope 5444D, Pico Technology). 
The diagnostic process was performed using the following pulse configurations: (1) standard HiPIMS, (2) standard bipolar HiPIMS, (3) chopped HiPIMS with five consecutive negative pulses, (4) bipolar HiPIMS with five cycles of alternating negative and positive pulses, (5) bipolar HiPIMS with five negative pulses, each followed by six short positive pulses, and (6) bipolar HiPIMS with five negative pulses, each followed by twelve short positive pulses (see \fref{fig:waveforms} for the waveforms and a shorthand notation used throughout the paper). During all HiPIMS conditions, the total duration of negative and positive pulses was kept constant separately, ensuring the same duty cycle for all configurations.

The Ar pressure was maintained at 1 Pa, with an Ar flow rate of 40 sccm. The average power delivered during the negative pulses was kept constant at 500 W. A positive voltage of 100 V was consistently applied during all positive pulses.

A passive thermal probe was used to determine the total energy flux from the plasma to the substrate during the deposition process. The probe was made from a stainless steel sheet (10 mm in diameter and 0.1 mm thickness), to which a thermocouple for temperature monitoring and an additional wire for applying voltage bias were welded from the back-side (see~\fref{fig:system}). The connecting wires were enclosed in ceramic tubes, providing thermal and electrical isolation from external influences. This probe operates by detecting the rate of probe temperature change ($\mathrm{d} \Tp$/$\mathrm{d}t$) during heating (when the plasma is on) and cooling (when the plasma is off) \cite{Kersten2000,Lundin2009a}. By monitoring these temperature variations, we can calculate the energy influx $Q_\mathrm{in}$ using the relationship:
\begin{equation}
Q_\mathrm{in} = C_\mathrm{p} \left[ \left( \ddt{\Tp} \right)_\mathrm{heat} - \left( \ddt{\Tp} \right)_\mathrm{cool} \right],
\label{eq:Qin}
\end{equation}
where $C_\mathrm{p}$ is the heat capacity of the probe. The probe was heated by the discharge up to $100\,\degC$ for each measurement. Afterwards, the plasma was turned off, and the cooling phase of the probe was recorded. An exponential decay equation was fitted to the temperature evolution in each phase. From these fits, the two terms in \eref{eq:Qin} were evaluated at the maximum temperature point. Since the probe heat capacity is unknown, the analysis focused on comparing the relative changes in total energy flux across different pulse configurations. No significant drift in $(\mathrm{d} T_\mathrm{p} / \mathrm{d} t)_\mathrm{cool}$ was observed across all measurements, indicating that the heat capacity of the probe remained consistent during the experiments. The standard deviation of $(\mathrm{d} T_\mathrm{p} / \mathrm{d} t)_\mathrm{cool}$ obtained from all measurements (36 in total) was used to determine the absolute error of the energy fluxes.

During a film deposition, the capacitance of the insulating surface (e.g., film, substrate, substrate holder) can differ\cite{REZEK2023}. To comprehensively examine the range of possible configurations, several conditions were investigated: (a) the thermal probe was left electrically floating, (b) it was connected to the ground through a capacitor (with a value of 1, 10 and 100 nF), as inspired by Du et al. \cite{Du2021}, (c) it was grounded via a diode, and (d) it was directly connected to the ground.

Figure \ref{fig:waveforms} shows the temporal evolution of the floating potential of the thermal probe positioned at the substrate distance (10 cm from the target surface) for various pulse configurations, along with the target voltage and current. Although the plasma potential was not directly measured here, previous studies indicate that it closely follows the target voltage trend, albeit at a slightly lower magnitude\cite{Batkova2020}. Also, as shown previously, the plasma potential is nearly uniform across the distance from the target to the substrate during the positive pulse for the same magnetron and magnetic field\cite{Pajdarova}. Consequently, the difference between the target voltage and the probe potential determines the magnitude of ion acceleration in the sheath around the probe.

As demonstrated in figure \ref{fig:waveforms}, when a positive pulse is applied in standard bipolar HiPIMS ([N+P]) compared to standard HiPIMS ([N]), ions may experience a slight acceleration at the onset of the positive pulse. However, this effect is very short-lived due to the charging of the substrate by the ions. In configuration 5[N+P], where the bipolar HiPIMS pulse is divided into smaller segments, ions can be accelerated at the beginning of each positive pulse before substrate charging restricts them, potentially providing more energy gain than for the standard [N+P] configuration. Following this idea, the configurations 5[N+6P] and 5[N+12P] use several very short positive pulses to increase the time for ion acceleration and allow the substrate to be discharged by electrons during the pauses. Note that the length of the off-times between the pulses ($\approx 4\,\mathrm{\mu s}$) was tuned so that the target potential drops to zero. However, it can be seen that the floating potential follows the target voltage practically instantly due to the very low capacitance of the thermal probe and its connecting wires ($200\,\mathrm{pF}$).

\begin{figure*}[t]
    \centering
    \includegraphics[width=1\linewidth]{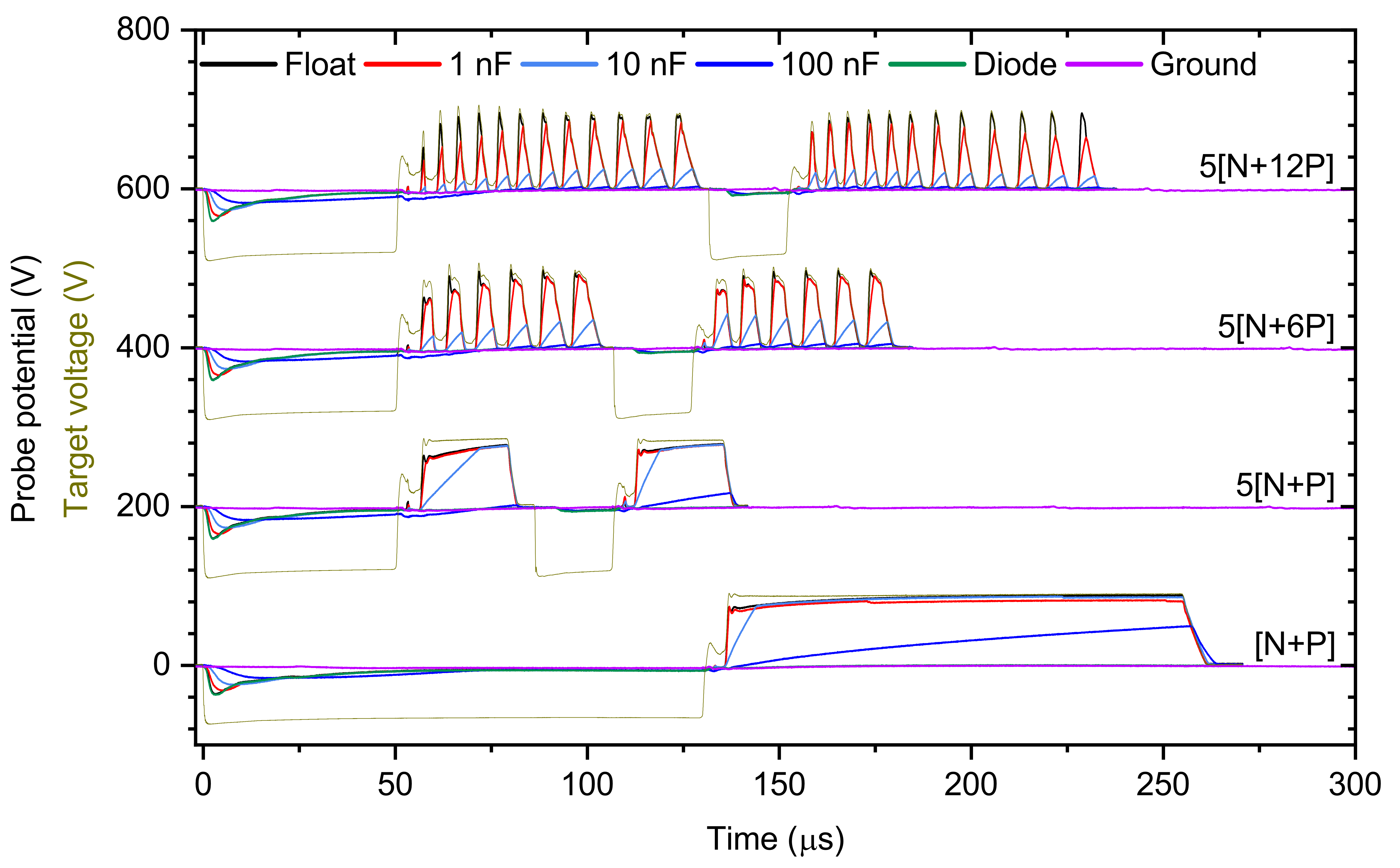}
       \caption{Waveforms of the floating potential of the thermal probe and target voltage for the investigated pulse configurations and conditions. It is important to note that the target voltage is reduced by a factor of 10 during negative pulses to enable a compact presentation of the waveforms while maintaining good resolution of the voltages during the positive pulses.}
    \label{fig:probpot}
\end{figure*}

The effect of the probe connection and its capacitance to ground is demonstrated in figure \ref{fig:probpot}, showing the probe potential and target voltage for different bipolar pulse configurations. The floating and the 1 nF capacitor case show similar behaviour, both characterized by quick charging, as evidenced by the quick rise of the measured voltage to a steady-state value near the applied target voltage. However, as larger capacitors are introduced, causing slower probe charge times, the difference between the probe potential and target voltage becomes more prominent. With a 10 nF capacitor, the probe potential stabilizes at a plateau $\approx 10 \,\mathrm{\mu s}$ later than for conditions with no capacitor in [N+P] configuration. In the 5[N+6P] and 5[N+12P] configurations, due to the short positive pulse duration, the probe potential does not ever rise close to the target voltage, which leads to a much larger average potential difference between the plasma and the probe. With the 100 nF capacitor, the substrate charges even more slowly (in $\approx 100 \,\mathrm{\mu s}$), allowing ion acceleration even for the long positive pulse (see [N+P] configuration). Additional chopping of the positive pulse (5[N+6P]) brings the probe potential closer to the ground, maintaining the high voltage difference between the plasma and the probe. Like in the 10 nF case, a variation in the number of short positive pulses ([5[N+12P]) does not significantly change the situation, as the probe potential is close to the ground anyway. If the probe is connected to the ground via a diode, it functions as an open circuit during the negative pulses, allowing the electrons to drive the probe to the negative floating potential. In contrast, it allows the discharging of the probe when its potential goes positive, effectively ensuring it remains grounded during the positive pulses. Under these conditions, the influx of electrons during the negative pulses is limited, as in the case of the floating probe, but the voltage difference between the probe and the plasma during the positive pulses is maximal.

\begin{figure}
    \centering
    \includegraphics[width=85mm]{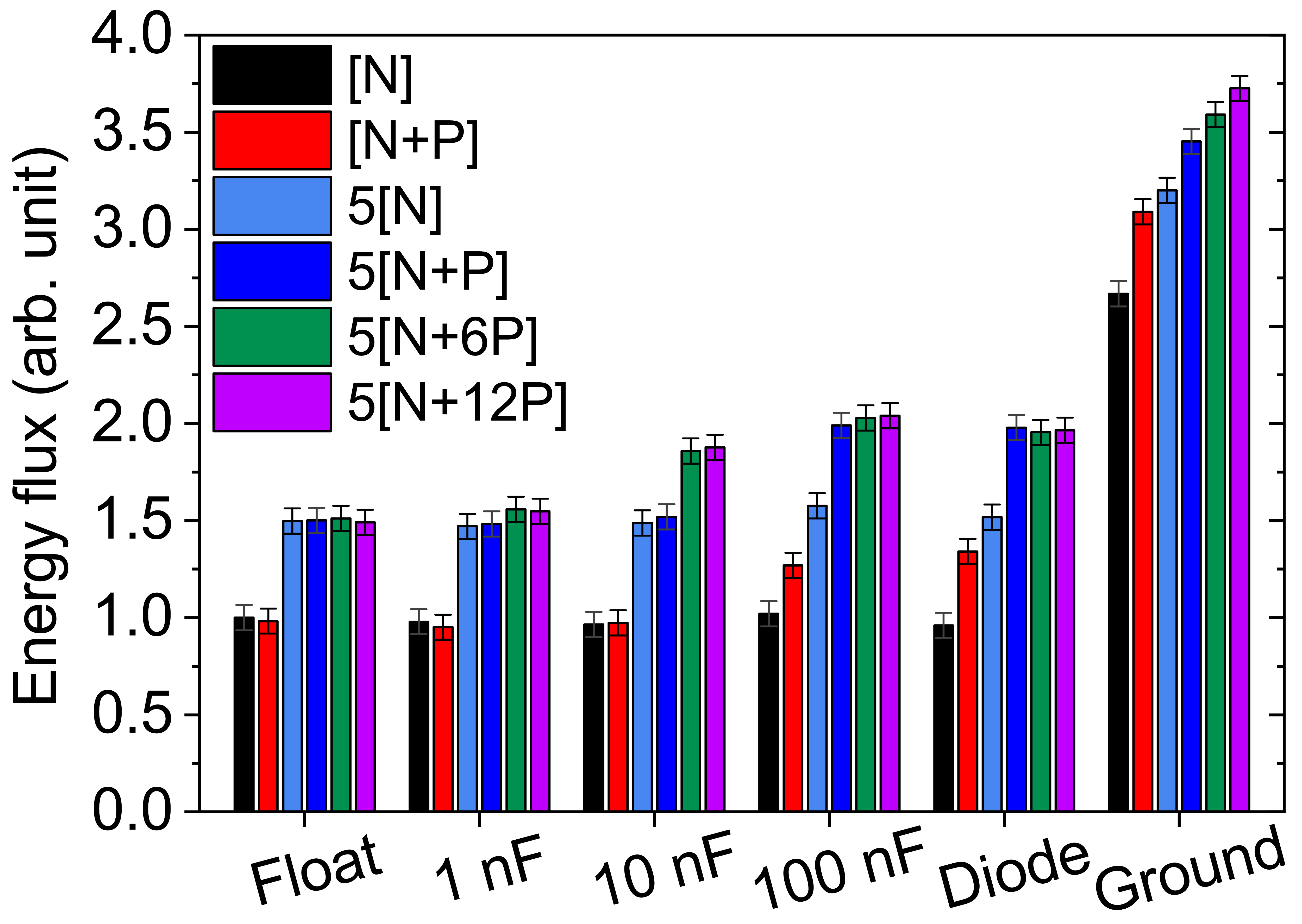}
   \caption{Normalized energy flux for different pulse configurations and conditions relative to standard HiPIMS for the probe at floating potential.}
    \label{fig:heatflux}
\end{figure}

Finally, the effect of these pulse configurations for different probe connections on the total energy influx from the plasma will be discussed. Note that the total energy influx measured by the thermal probe, $Q_\mathrm{in}$, is the sum of several contributions: $Q_\mathrm{c}$, the heat of condensation from the deposited film; $Q_\mathrm {i}$, the energy contribution from ions; $Q_\mathrm {e}$, the energy contribution from electrons; and $Q_\mathrm{rec}$, the energy released from ion-electron recombination \cite{Kersten2000}. Figure \ref{fig:heatflux} shows the measured energy flux normalized to the energy flux obtained for standard HiPIMS with a floating probe.

In the floating probe case, there is no significant difference between HiPIMS [N] and bipolar HiPIMS [N+P], confirming that the probe quickly charges during the positive pulse by the incident ions, as seen in \fref{fig:probpot}. The possible higher voltage difference between the plasma and the probe at the beginning of the positive pulse does not last long enough to provide a significant difference in the energy flux (concerning the measurement error). However, there is a $50\,\%$ increase in energy flux for the other configurations with 5 negative pulses, including the chopped HiPIMS configuration 5[N] with no positive pulses. Finally, figure \ref{fig:heatflux} shows that the addition of one or more positive pulses after each negative pulse (5[N+P], 5[N+6P] and 5[N+12P] configurations) does not provide any increase in energy flux when the probe is floating. This is again because of the fast charging of the probe.

It is worthwhile to discuss the $50\,\%$ increase in energy flux between the [N] and 5[N] configurations in more detail. This is in agreement with Barker et al.\cite{Barker2014}, who observed (for titanium) a simultaneous increase in the deposition rate and the ionized fraction estimated at the substrate position by chopping the HiPIMS pulse. In our study, the deposition rate increased by only 20\%, implying that a corresponding increase in ion contribution is necessary to explain the observed enhancement in energy flux. The increase in the deposition rate can be attributed to the reduction of self-sputtering (a process with a lower sputtering yield than sputtering by argon ions), as self-sputtering cannot fully develop during short pulses\cite{Barker2013}. Additionally, the self-sputtering reduction is supported by lower gas rarefaction facilitated by partial refilling of the argon gas near the target during the off-times between pulses \cite{Lundin2009}. However, this effect is likely less pronounced here due to the relatively short off-time used. The increase in the ion flux towards the probe can be ascribed to the increased average ionization degree of atoms in the plasma, as reported by Barker et al.\cite{Barker2014}. Note that the target voltage during the negative pulses is much higher for 5[N] than for [N]. Also, the peak current in the first two pulses of 5[N] is larger than the peak current in [N]. On the other hand, Hnilica et al.\cite{Hnilica2023} reported a lower Ti ionization degree for two consecutive pulses compared to one pulse since, in their case, the peak current in the two pulses was much lower than the peak current in the one-pulse configuration (the duty cycle was not fixed). It is also argued that multiple shorter pulses effectively lower the ion back-attraction to the target by allowing more ions to reach the substrate at the end of each pulse\cite{Antonin2015, Rudolph2020}. Thus, a larger fraction of ions generated during the pulse reaches the substrate. It should be mentioned that this study cannot distinguish the contributions of Ar$^+$ and Ti$^+$ ions to the energy flux. In the case of the above-mentioned reduced self-sputtering, leading to a lower fraction of Ti$^+$ ions above the target, it is likely that the Ar$^+$/Ti$^+$ ion fraction at the substrate increases for the 5[N] configuration compared to the [N] configuration.

When a 1 nF capacitor is connected between the probe and the ground, the charging of the probe is slowed down. In \fref{fig:probpot}, this is barely noticeable, but \fref{fig:heatflux} reveals a small increase in energy flux for the regimes with many short positive pulses (5[N+6P] and 5[N+12P]), compared to the floating probe case. This trend is even more exaggerated for the 10 nF capacitor. As the probe charging time scale becomes comparable to the length of the positive pulses in the 5[N+6P] and 5[N+12P] configurations, a significant ($20\,\%$) increase in the energy flux is observed compared to the 5[N] and 5[N+P] cases. Note that there is no significant difference among the 5[N+6P] and 5[N+12P] configurations. The primary reason is that the average probe potential is comparable, and the total length of the positive pulses is equal (by the design of this study). Nevertheless, the plasma density after a negative pulse decays in time, so the precise timing of the sequence of positive pulses with respect to the preceding negative pulses can become significant. For the 100 nF capacitor, the trend slightly changes. First, there is a difference between the [N] and [N+P] configurations, as the charge time of the probe is so slow that even the long positive pulse becomes effective for ion acceleration. For the same reason, the 5[N+P] configuration exhibits much higher energy flux than the 5[N] configuration. The 5[N+6P] and 5[N+12P] configurations are now comparable to the 5[N+P] configuration. Since the probe potential is close to the ground for this high capacitor, see \fref{fig:probpot}, chopping the positive pulses brings no additional benefit over the longer positive pulses.

The conditions employing the diode closely resemble the trend obtained for the 100 nF capacitor. For the grounded probe, compared to the other conditions, the energy contributions from $Q_\mathrm {e}$ increase significantly, as the continuous ground potential connection facilitates increased fluxes of electrons to the probe during the negative pulses. As a result, it leads to the highest energy flux across all pulse configurations, particularly in setups with a frequently chopped positive pulse.

From the results presented above, it is evident that the effectiveness of the chopped pulsed configuration to increase the energy flux to films strongly depends on the capacitance of the surface. For a conducting substrate on an insulated substrate holder or an insulating substrate (like standard laboratory glass), the capacitance of the surface will typically be very low (below $1\,\mathrm{nF}$). In such cases, our results indicate that bipolar HiPIMS does not effectively increase the energy delivered to the film due to rapid charging of the surface. This is in agreement with our recent work where pulsed DC bias of the substrate holder provided no effect on the structure of ZrO$_2$ films deposited on glass\cite{REZEK2023}. In principle, a capacitance of the insulated surface on the order of $10\,\mathrm{nF}$ could be attained, for example, when a thin insulating film is deposited on a conducting substrate\cite{Du2021, REZEK2023}. In such a case, bipolar HiPIMS with a chopped positive pulse (configurations 5[N+6P] and 5[N+12P]) will effectively increase the energy delivered to the film by plasma ions.

Note that the capacitance values presented in this work are not universally transferable to other deposition conditions. The relevant parameter for evaluating the effectiveness of chopped bipolar pulse configurations is the time necessary to charge the insulated surface 
\begin{equation}
\tau \approx \frac{C U_\mathrm{rev}}{J A_\mathrm{s}} \,,
\end{equation}
where $C$ is the surface capacitance, $U_\mathrm{rev}$ is the positive target voltage, $J$ is the ion current density to the surface and $A_\mathrm{s}$ is the surface area. The length of the positive pulses should be smaller than $\tau$ for effective ion acceleration.

To conclude, this study explores the optimization of bipolar High-power Impulse Magnetron Sputtering (HiPIMS) pulse configurations to enhance energy delivered to insulating surfaces by plasma ions. Various pulse configurations, including unipolar, bipolar, chopped unipolar, and chopped bipolar HiPIMS, were studied, revealing that chopped pulse configurations can enhance energy flux to the substrate under certain conditions. Our study confirmed prior findings that employing multiple (chopped) negative pulses substantially enhances the energy flux to the substrate relative to a single HiPIMS pulse. Additionally, in the case of bipolar HiPIMS, the energy flux was observed to be influenced by the capacitance of the insulated surface. At low capacitance, increased ion acceleration occurs only at the beginning of each positive pulse before the surface becomes charged to the floating potential. For capacitances below $1\,\mathrm{nF}$, applying the positive pulse does not significantly enhance the energy flux. In contrast, high capacitance allows for a large potential difference between plasma and surface throughout the pulse, even for long positive pulses, significantly improving energy delivery. For a medium capacitance (in our case around $10~\,\mathrm{nF}$), chopping the positive pulse into several smaller pulses can achieve increased energy fluxes to the surface, almost as high as with the case of the grounded surface (when electron heating is suppressed). This is facilitated by the fact that the time of surface charging is comparable to the length of short (chopped) positive pulses.

\section*{Acknowledgments}
This work was supported by the project Quantum materials for applications in sustainable technologies (QM4ST), funded as project No. CZ.02.01.01/00/22\_008/0004572 by Programme Johannes Amos Commenius, call Excellent Research. 

\section*{References}
\bibliographystyle{iopart-num}
\bibliography{references}

\end{document}